\documentclass[11pt]{article}

\usepackage{microtype} 
\usepackage{booktabs}  
\usepackage{url}  

\usepackage{amsmath}
\usepackage{amsthm}

\usepackage{graphicx}

\usepackage[final]{latex-simple}
\usepackage[final]{latex-simple}

\usepackage{natbib}

\title{Multi-variable Hard Physical Constraints for Climate Model Downscaling}

\author[1]{\nameemail{Jose González-Abad}{gonzabad@ifca.unican.es}}
\author[2,3]{\nameemail{Álex Hernández-García}{}}
\author[4,5]{\nameemail{Paula Harder}{}}
\author[2,6]{\nameemail{David Rolnick}{}}
\author[1]{\nameemail{José Manuel Gutiérrez}{}}

\affil[1]{Instituto de Física de Cantabria (IFCA), CSIC-UC}
\affil[2]{Mila Quebec AI Institute}
\affil[3]{University of Montreal}
\affil[4]{Fraunhofer ITWM}
\affil[5]{University of Kaiserslautern}
\affil[6]{McGill University}

\hypersetup{%
  pdfauthor={}, 
  pdftitle={},
  pdfsubject={},
  pdfkeywords={}
}

\begin{document}

\maketitle

\begin{abstract}
Global Climate Models (GCMs) are the primary tool to simulate climate evolution and assess the impacts of climate change. However, they often operate at a coarse spatial resolution that limits their accuracy in reproducing local-scale phenomena. Statistical downscaling methods leveraging deep learning offer a solution to this problem by approximating local-scale climate fields from coarse variables, thus enabling regional GCM projections. Typically, climate fields of different variables of interest are downscaled independently, resulting in violations of fundamental physical properties across interconnected variables. This study investigates the scope of this problem and, through an application on temperature, lays the foundation for a framework introducing multi-variable hard constraints that guarantees physical relationships between groups of downscaled climate variables.
\end{abstract}

\section{Introduction}
\label{Introduction}

Global Climate Models (GCMs) are physics-based models employed to simulate the spatio-temporal evolution of climate and to obtain future climate projections under different climate change scenarios. The resulting projections are crucial to develop adaptation and mitigation plans in many sectors. However, due to computational and physical limitations, the resolution of these models is coarse, which hinders their use in regional-to-local applications.

Statistical Downscaling (SD) techniques attempt to overcome this limitation by learning a relationship between large-scale (low-resolution) data and local-scale (high-resolution) variables of interest. The so-called Perfect Prognosis approach (PP-SD) \citep{maraun2018statistical} aims at learning a relationship between large-scale predictors and local-scale predictand time-matched pairs from observational data. The set of predictors describes the state of the atmosphere, whereas the predictand corresponds to a surface variable of interest such as temperature or precipitation. 

Deep Learning (DL) has recently emerged as a promising SD method, with great potential given its ability to handle spatio-temporal data and model non-linear relationships. DL models developed using historical and future GCM projections can produce actionable local-scale downscaled predictions for climate change studies \citep{bano2022downscaling}. 

Climate is a highly complex system and the variables involved in climate models are closely related to each other by physical links imposed by the constitutive equations, which take into account the interactions and feedback among them. However, most previous works in SD have proposed methods that downscale variables independently, ignoring potential relationships. This can result in inconsistencies and violations of basic physical properties between related groups of variables. Such violations raise important concerns about the reliability and adoption of DL-based downscaling models in climate change applications.

In this work, we first examine the violations of basic physical properties introduced by DL models for the PP-SD of temperature. Our analysis reveals that physical constraints are largely violated when variables are downscaled independently (univariate downscaling), which is especially pronounced in the attempt to generalize from present conditions to future climate scenarios. To address this limitation, we investigate the potential benefits of using a shared model to simultaneously downscale a group of variables. Going further, we establish the groundwork of a novel framework to incorporate multi-variable, physical, hard constraints in neural networks. Through our experimental setup with two architectures and evaluations in GCM projections of future climate, we demonstrate that our framework guarantees the physical constraints for the downscaling of temperature and can be flexibly adapted to other architectures. Our proposed multi-variable model strengthens the reliability of DL-based downscaling and hence facilitates its adoption by the climate science community for practical applications.

\section{Background}
\label{Background}

\textbf{Deep learning methods for univariate downscaling.}
Many works have explored the use of DL models for downscaling individual variables, most using the so-called super-resolution approach \citep{vandal2017deepsd,stengel2020adversarial,wang2021fast,kumar2021deep,passarella2022reconstructing,sharma2022resdeepd}. Inspired by super-resolution methods in computer vision \citep{wang2020deep}, these techniques use a low-resolution version of the target variable as a predictor (in contrast to PP, where the input and output represent different physical quantities). However, this approach is less suitable for downscaling GCMs, as the surface variables used as input are unreliable predictors due to the coarse resolution at which GCMs operate. 

Deep learning approaches for univariate PP-SD have been proposed by \citet{pan2019improving,bano2020configuration,sun2021statistical,quesada2022repeatable,rampal2022high}. These models rely on large-scale atmospheric variables as predictors, which are better reproduced by GCMs since these do not depend on local-scale dynamics. These models show promising results in projecting plausible future climate change scenarios over Europe based on different GCMs \citep{bano2021suitability,bano2022downscaling}.

\textbf{Approaches for multivariate downscaling.} Non-DL-based SD models operating on multiple interrelated variables have been extensively explored in the literature. Typically, multivariate downscaling enhances multivariate properties such as cross-correlation \citep{jeong2012multivariate,khalili2013statistical,eum2020effects}, although individual variables may be negatively impacted \citep{bhowmik2017multivariate}. Multivariate DL-based SD approaches, by contrast, have not yet been well developed. \citet{wang2022deep} trained a multivariate DL model for downscaling minimum and maximum temperatures, but found that some predicted minimum temperatures were higher than the corresponding maximum temperatures, thus violating basic physical properties. No previous work has specifically addressed such physical inconsistencies. 

\textbf{Constrained DL in climate modeling.} Several works have attempted to enforce physical constraints in DL models for climate and weather, through both soft and hard constraints. Soft constraints are introduced by adding additional loss terms to the model \citep{esmaeilzadeh2020meshfreeflownet,beucler2021enforcing, harder2022physics}. Hard constraints are implemented by introducing modifications in the neural network architecture \citep{geiss2020strict,harder2022physics,hess2022physically}, ensuring that the constraints are satisfied during learning and inference. In \citet{harder2022generating,geiss2022downscaling}, hard constraints are applied to DL models for super-resolution SD, enforcing physical relationships between the high-resolution predictand and its low-resolution predictor.

\section{Experimental framework}
\label{Experimental}

\subsection{Region of study and data}
\label{Data}

\begin{figure*}
    \centering
    \includegraphics[width=\columnwidth]{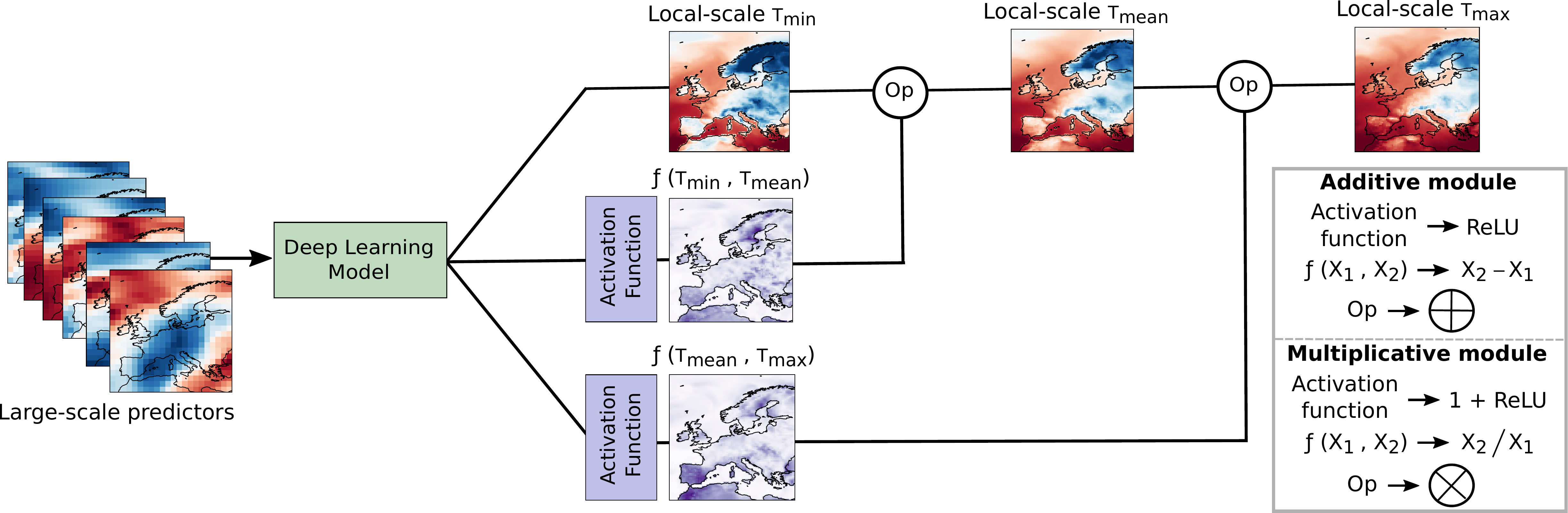}
    \caption{Schematic view of the proposed additive and multiplicative modules. The general architectural layout is depicted in the main area, while the specific operations and activation functions of the modules are indicated on the right side.}
    \label{fig:module-schema}
\end{figure*}

Our experimental framework is based on the PP approach. Following previous work \citep{bano2020configuration}, in order to represent the atmospheric state, we choose five large-scale variables (geopotential height, zonal and meridional wind, air temperature, and specific humidity) at four different vertical levels (1000, 850, 700 and 500 hPa) as predictors. As predictand, we focus on three variables, namely, minimum, mean, and maximum near-surface air temperature. The predictor variables are obtained from the ERA-Interim reanalysis data set \citep{dee2011era} at a 2$^\circ$ spatial resolution, while the predictand variables are extracted from the W5E5 observational data set \citep{lange2019wfde5} at a 0.5$^\circ$ resolution, both at a daily timescale. We choose the European continent as the spatial domain to conduct our experiments.

We select the EC-Earth model run \textit{r12i1p1} \citep{doblas2018using} as the GCM to downscale. We focus on the period of 2006-2100 of the Representative Concentration Pathway 8.5 scenario (RCP8.5) \citep{schwalm2020rcp8}. This scenario is selected as it represents the strongest climate change signal among those developed in the Coupled Model Intercomparison Project Phase 5 (CMIP5, \citet{taylor2012overview}).

\subsection{Deep learning downscaling models}
\label{Models}
We train two different DL models: UNet \citep{ronneberger2015u} and DeepESD \citep{bano2022downscaling}. UNet is a fully-convolutional model inspired by an architecture widely used in image segmentation, which has also demonstrated outstanding performance in climate applications, including statistical downscaling \citep{quesada2022repeatable,wang2022deep,doury2023regional}. DeepESD is a model developed for the PP-SD of temperature and precipitation over Europe. It is composed of a set of convolutional layers and a final fully-connected layer. This model has been validated for the downscaling of various GCMs under future climate scenarios \citep{bano2021suitability,bano2022downscaling}. We choose these two distinct models as reflecting the current state-of-the-art DL models for downscaling.

The models are trained using the ERA-Interim and W5E5 data sets as predictor and predictand fields, respectively. Predictors are standardized to conform to a standard normal distribution, while predictands are normalized to take on values within the interval $[0, 1]$. We divide the observational data into a training (1980-2000) and a test set (2001-2005). Models are trained to minimize the Mean Squared Error (MSE) using the Adam optimizer with a standard learning rate of $10^{-5}$ and batch size of 64, using early stopping to prevent overfitting. Prior to passing GCM predictors to DL models for downscaling future scenarios, we perform a signal-preserving adjustment of the monthly mean and variance of the GCM, as suggested by \citet{bano2022downscaling}. This improves the extrapolation capabilities of the models by allowing for better similarity in the distributions of the GCM and the reanalysis data.

\subsection{Multi-variable hard physical constraints}
\label{Hard-Constraints}
In this work, we introduce methods for enforcing hard physical constraints between interrelated variables in SD, in the form of simple modules that can be appended onto DL downscaling architectures. 

The predictands used in our study are the near-surface minimum, mean, and maximum temperature. For these variables, the following constraints obviously hold:
\begin{equation}
    T_{min} \leq T_{mean} \leq T_{max}
    \label{eq:constraints}
\end{equation}
where $T_{min}$, $T_{mean}$, and $T_{max}$ represent the minimum, mean, and maximum temperature. Any violation of these constraints would diminish not merely the average accuracy of an SD method but also its plausibility, significantly reducing the likelihood that the method would be used in practice. While prior work \citep{wang2022deep} predicted $T_{min}$ and $T_{max}$ jointly, they still suffered from significant constraint violation. We propose two different options for predicting these three variables together while enforcing physical constraints. Figure \ref{fig:module-schema} presents a schematic illustration of our proposed additive and multiplicative modules.
 
In the \textit{additive} approach, we predict $T_{min}$ and $T_{mean}-T_{min}$ and add these values to obtain $T_{mean}$, likewise computing $T_{max}$ by predicting $T_{max}-T_{mean}$ and adding it to the previously estimate for $T_{mean}$. This method ensures that all constraints are met by using \textit{ReLUs} to enforce the nonnegativity of the predicted $T_{min}$, $T_{mean}-T_{min}$, and $T_{max}-T_{mean}$. In the \textit{multiplicative} approach, a similar method is applied, but instead, we predict $T_{min}$ using \textit{ReLU} activation and the values $T_{mean}/T_{min}$ and $T_{max}/T_{mean}$ using the activation function $1+\text{\textit{ReLU}}$. Multiplying these predicted values gives the desired quantities while ensuring constraint satisfaction.

In order to assess the efficacy of the proposed modules, we conducted four experiments. In accordance with established methodologies, the first experiment (referred to as \textit{single}) entails the independent modeling of each variable through a distinct DL model for each one. The second experiment (referred to as \textit{shared}) involves training a shared model for the three variables in a multivariate approach. In the case of the shared UNet, the sole discrepancy from the single model is the last convolutional layer, which now features three output channels rather than one. In the DeepESD model, the shared approach entails the utilization of three parallel fully-connected layers instead of one. In the third (\textit{additive}) and fourth (\textit{multiplicative}) experiments, multi-variable hard-constrained models are trained with the two approaches presented above.

\section{Results}
\label{Results}

\begin{figure}
    \vskip 0.2in
    \begin{center}
    \centerline{\includegraphics[width=0.8\columnwidth]{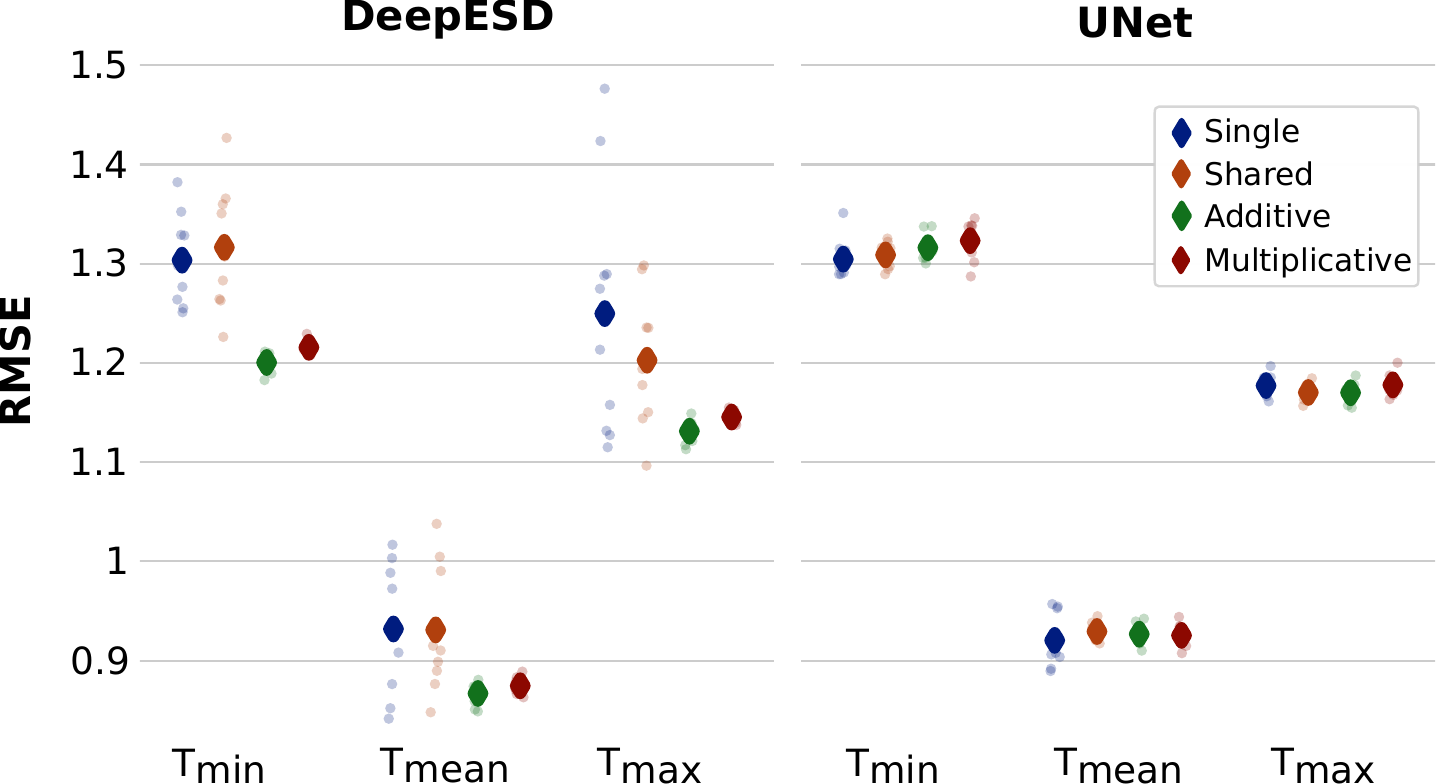}}
    \caption{RMSE on the test set for the four experiments (\textit{single}, \textit{shared}, \textit{additive} and \textit{multiplicative}) of the two different models (UNet and DeepESD) trained to downscale minimum ($T_{min}$), mean ($T_{mean}$) and maximum ($T_{max}$) temperature. The values for 10 independent executions and their means are shown.}
    \label{fig:RMSE}
    \end{center}
\end{figure}

Figure \ref{fig:RMSE} shows the Root Mean Squared Error (RMSE) values on the test set obtained for the two DL models trained, namely UNet and DeepESD, for the different experiments (\textit{single}, \textit{shared}, \textit{additive} and \textit{multiplicative}). For $T_{min}$ and $T_{max}$ all combinations of models and experiments display a higher error than for $T_{mean}$. The underlying cause for this is the extreme values associated with $T_{min}$ and $T_{max}$, which make modelling them harder \citep{kjellstrom2007modelling}. 

The UNet model exhibits comparable RMSE values across all experiments. However, for the DeepESD model, a notable improvement in RMSE performance occurs for both the additive and multiplicative hard-constrained models as compared to the single and shared versions. This improvement may be attributed to the presence of fully-connected layers in the architecture of DeepESD, which give rise to the phenomenon known as \textit{dying ReLUs} \citep{lu2019dying}, thereby resulting in a degradation of the model's predictive performance. In the hard-constrained modules, the \textit{ReLU} activation outputs are incorporated into $T_{min}$ and $T_{mean}$ through addition or multiplication, thereby mitigating the possible adverse effects of this phenomenon.

\begin{figure}[ht]
    \vskip 0.2in
    \begin{center}
    \centerline{\includegraphics[width=0.6\columnwidth]{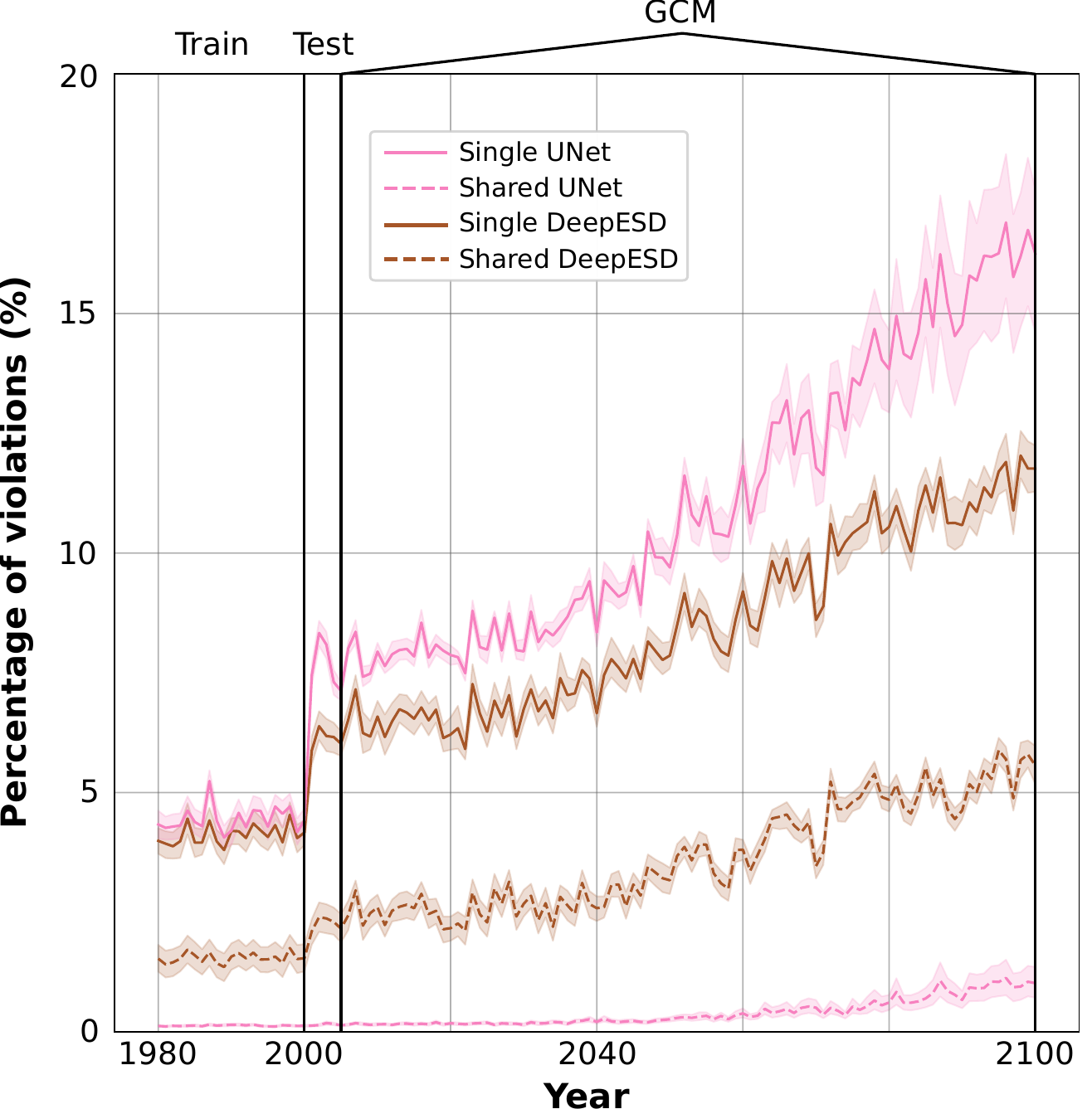}}
    \caption{Annual percentage of violations of the physical constraints for the two different DL models intercompared (UNet and DeepESD), for the \textit{single} and \textit{shared} experiments. For each series, the $95\%$ confidence interval is shown.}
    \label{fig:number-violations}
    \end{center}
\end{figure}

Figure \ref{fig:number-violations} presents the annual percentage of violations of the constraints introduced in Section \ref{Hard-Constraints} for the downscaled variables obtained from the UNet and DeepESD models. For each of these, the figure displays the results for the \textit{single} and \textit{shared} experiments (solid and dashed lines, respectively), along with the $95\%$ confidence interval computed by applying bootstrapping to 10 independent executions. Violations for the \textit{additive} and \textit{multiplicative} experiments are not shown since, by construction, they are zero.

In the \textit{single} experiments, both architectures show a similar trend in the amount of violations for the training period (1980-2000), maintaining a relatively low and constant amount of violations. However, during the test period (2001-2005), the models must extrapolate to out-of-distribution data, resulting in a sudden spike in the percentage of violations. As the models are applied to the GCM predictors (2006-2100), this issue becomes more pronounced, particularly as we move forward in time, reflecting the difficulties encountered while dealing with the extreme climate conditions associated with the RCP8.5 scenario. The confidence interval of the DeepESD model remains stable over time; however, UNet variability increases, thus being more susceptible to these extreme conditions.

In the \textit{shared} experiment, UNet demonstrates a marked reduction in the percentage of violations, whereas DeepESD continues to exhibit a significant amount. This disparity is attributed to the architectural composition of DeepESD, which allocates a substantial portion of its parameters in the final fully-connected layer. Consequently, the shared model replicates this layer across the downscaled variables, with each variable retaining a considerable amount of parameters dedicated solely to itself, thus not being too different from the single version. Unlike DeepESD, UNet consists entirely of convolutional layers; therefore, in its shared version, each variable's allocation of parameters remains minimal. Despite the evident decline in violations in the \textit{shared} experiments, they do not completely disappear, and more importantly, there are no guarantees regarding the behavior of these models under diverse future scenarios and various GCMs. What is more, as in the \textit{single} experiment, the variability in the number of violations of the UNet model increases with time. Only the multi-variable hard physically constrained models can assure no violations in any scenario.

\section{Conclusions}
\label{Conclusions}

In this work, we have analyzed the violations of physical constraints commited by DL models for the PP-SD of temperature. Our results reveal that current approaches result in a large number of violations, particularly in the GCM domain. To address this, we have proposed a shared model for the desired variables as a partial solution. However, to ensure multi-variable physical constraints, we have introduced a simple and flexible framework that satisfies hard constraints and achieves the same performance or better than standard approaches.

Preserving fundamental physical properties of GCMs in DL models is crucial for their reliability and hence the adoption by the climate science community. For instance, in the Sixth IPCC Assessment Report, certain indices relying on daily minimum, mean and maximum temperature are employed to asses the impact of climate change in energy demand \citep{IPCC_2023}. In this study, we set the groundwork for a novel framework that enables DL models to mimic some underlying physics properties of GCMs, enhancing the reliability of PP-SD models. This work represents a significant effort in integrating DL and GCM models. In future research, we plan to extend our approach to new variables and constraints, further enhancing the performance and reliability of DL models for PP-SD. 

\begin{acknowledgements}
J. González-Abad would like to acknowledge the support of the funding from the Spanish Agencia Estatal de Investigación through the Unidad de Excelencia María de Maeztu with reference MDM-2017-0765.
\end{acknowledgements}


\bibliography{references}
\bibliographystyle{latex-simple}




\end{document}